\documentclass[prd,aps,showpacs,epsf,floats,10pt]{revtex4}%
\usepackage{amssymb}
\usepackage{amsfonts}
\usepackage{amsmath}
\usepackage{graphicx}%
\providecommand{\U}[1]{\protect\rule{.1in}{.1in}}
\begin{document}
\title{SOFTENING THE COMPLEXITY OF ENTROPIC MOTION ON CURVED STATISTICAL MANIFOLDS}
\author{Carlo Cafaro$^{1}$, Adom Giffin$^{2}$, Cosmo Lupo$^{3}$, Stefano Mancini$^{4}$}
\affiliation{$^{1\text{, }3\text{, }4}$School of Science and Technology, Physics Division,
University of Camerino, I-62032 Camerino, Italy}
\affiliation{$^{2}$Princeton Institute for the Science and Technology of Materials,
Princeton University, Princeton, NJ 08540, USA}

\begin{abstract}
We study the information geometry and the entropic dynamics of a $3D$ Gaussian
statistical model. We then compare our analysis to that of a $2D$ Gaussian
statistical model obtained from the higher-dimensional model via introduction
of an additional information constraint that resembles the quantum mechanical
canonical minimum uncertainty relation. We show that the chaoticity (temporal
complexity) of the $2D$ Gaussian statistical model, quantified by means of the
Information Geometric Entropy (IGE) and the Jacobi vector field intensity, is
softened with respect to the chaoticity of the $3D$ Gaussian statistical model.

\end{abstract}

\maketitle

\section{Introduction}

A very important problem in modern science concerns the description and the
understanding of the elusive concept of complexity \cite{L88, GM95, F98}.
There are pragmatic reasons of primary importance in quantum information
science that justify the study of complexity, for example the problem of
quantifying how complex is quantum motion \cite{C09}. Unfortunately, our
knowledge of the basic connections between complexity, dynamical stability,
and chaoticity in a fully quantum domain is not satisfactory \cite{C09, O98}.
The concept of complexity is very difficult to define, its origin is not fully
understood \cite{W84, W85, R95, H86, G86, JPC89} and it is mainly for this
reason that several quantitative measures of complexity have appeared in the
scientific literature \cite{L88, GM95, F98}. In classical physics, complexity
measures are settled in a much better way. The Kolmogorov-Sinai metric entropy
\cite{K65}, that is the sum of all positive Lyapunov exponents \cite{P77}, is
a powerful indicator of unpredictability in classical systems and it measures
the algorithmic complexity of classical trajectories \cite{B83, B00, S89,
W78}. Other known measures of complexity are the logical depth \cite{B90}, the
thermodynamic depth \cite{S88}, the computational complexity \cite{P94}, the
stochastic complexity \cite{R86} and many more. Ideally, a good definition of
complexity should be mathematically rigorous and intuitive at the same time so
that we are able to tackle complexity-related problems in computation theory
and statistical physics as well. Of course, a quantitative measure of
complexity is truly useful if its range of applicability is not limited to few
unrealistic applications. For this reason, in order to properly define
complexity measures, the purpose for defining such a measure and what it is
intended to capture should be clearly stated.

Another problem of great theoretical interest is understanding how to compare
quantum and classical chaos (temporal complexity) and explaining the reason
why the former is weaker than the latter \cite{caron1, caron2, kroger}.
Indeed, it is commonly conjectured that the weakness of quantum chaos may be a
consequence of the Heisenberg uncertainty relation \cite{caron1, caron2,
kroger}. It is also known that a quantum description of chaos is qualitatively
different from a classical description and that the later cannot simply be
considered an approximation of the former. Indeed, the only trace of quantum
theory which a classical description may retain is the canonical Heisenberg's
uncertainty relation, namely a minimum spread of order $\hbar^{n}$ in the
$2n$-dimensional phase space \cite{peres} ($\hbar\overset{\text{def}}{=}%
\frac{h}{2\pi}$ where $h$ is Planck's constant).

Motivated by such considerations and following the line of reasoning presented
in \cite{cafaroPD}, we will discuss in this paper the manner in which the
degree of complexity changes for a statistical model (the probabilistic
description of a physical system) in the presence of incomplete knowledge
("softening effects") when the entropic dynamics (or, information-constrained
dynamics) \cite{catichaED} on the underlying curved statistical manifolds
becomes more constrained. By "softening effects" we mean any attenuation in
the asymptotic temporal growth of complexity indicators of motion. Our aim is
to reduce the probabilistic description of dynamical systems in the presence
of partial knowledge to information geometry (Riemannian geometry applied to
probability theory, see \cite{amari}) and inductive inference \cite{c0, c1,
c2, c3, c4}. To achieve such a task, we have developed a theoretical framework
termed IGAC, \emph{Information Geometric Approach to Chaos}
\cite{cafarothesis, cafaroCSF}, where information geometric techniques are
combined with Maximum relative Entropy methods \cite{c0, c1, c2, c3, c4} to
study the complexity of informational geodesic flows on curved statistical
manifolds (statistical models) underlying the probabilistic description of
physical systems in the presence of incomplete information. IGAC is the
information geometric analogue of conventional geometrodynamical approaches to
chaos \cite{casetti, di bari} where the classical configuration space is being
replaced by a statistical manifold with the additional possibility of
considering chaotic dynamics arising from non conformally flat metrics (the
Jacobi metric is always conformally flat, instead). It is an
information-geometric extension of the Jacobi geometrodynamics (the
geometrization of a Hamiltonian system by transforming it to a geodesic flow
\cite{jacobi}). The reformulation of dynamics in terms of a geodesic problem
allows the application of a wide range of well-known geometrical techniques in
the investigation of the solution space and properties of the equation of
motion. The power of the Jacobi reformulation is that all of the dynamical
information is collected into a single geometric object in which all the
available manifest symmetries are retained- the manifold on which geodesic
flow is induced. For example, integrability of the system is connected with
existence of Killing vectors and tensors on this manifold. The sensitive
dependence of trajectories on initial conditions, which is a key ingredient of
chaos, can be investigated from the equation of geodesic deviation (the
so-called Jacobi-Levi-Civita equation or JLC equation). Within the IGAC
formalism, conventional signatures of the chaoticity of classical dynamics
emerge via the asymptotic exponential divergence of the JLC vector field
intensity and/or via the specific behavior of the asymptotic growth of the
Information Geometric Entropy (IGE) (i.e., the logarithm geodesic volume)
\cite{cafaroPD}.

In view of the above-mentioned considerations, we discuss here the information
geometry and the information-constrained dynamics of a $3D$ Gaussian
statistical model. We then compare our analysis to that of a $2D$ Gaussian
statistical model obtained from the higher-dimensional model via introduction
of an additional information constraint that resembles the quantum mechanical
canonical minimum uncertainty relation. We show that the chaoticity (temporal
complexity) of the $2D$ Gaussian statistical model, quantified by means of the
IGE \cite{cafaroPD} and the Jacobi vector field intensity, is softened (that
is, attenuated) with respect to the chaoticity of the $3D$ Gaussian
statistical model. In view of the similarity between the information
constraint on the variances and the phase-space coarse-graining imposed by the
Heisenberg uncertainty relations, we suggest that our work provides a possible
way of explaining the phenomenon of suppression of classical chaos operated by quantization.

The layout of this article is as follows. In Section II\textbf{, }we present
the basic differential geometric properties of both the $3D$ and $2D$ Gaussian
statistical models. In\ Section III, we describe the geodesic paths on the
curved statistical manifolds underlying the entropic dynamics of the three and
two-dimensional statistical models. In Section IV, we study the chaotic
properties of the information-constrained dynamics on the underlying curved
statistical manifolds by means of the IGE and the Jacobi vector field
intensity. Our final remarks appear in Section V.

\section{Geometry of the Statistical Models}

The statistical models studied are a $3D$ uncorrelated Gaussian model and a
$2D$ Gaussian statistical model obtained from the higher-dimensional model via
introduction of an additional information constraint that resembles the
canonical minimum uncertainty relation in quantum theory. For a brief and
recent overview on the IGAC, we refer to \cite{cafaroPD}. Note that the
dimensionality\textbf{\ (}$2D$\textbf{, }$3D$\textbf{) }pertains to the
macroscopic variables.

\subsection{The 3D Statistical Model}

The probability distributions $p\left(  x\text{, }y|\mu_{x}\text{, }\sigma
_{x}\text{, }\sigma_{y}\right)  $ that characterize the $3D$ Gaussian
statistical model are given by,%
\begin{equation}
p\left(  x\text{, }y|\mu_{x}\text{, }\sigma_{x}\text{, }\sigma_{y}\right)
\overset{\text{def}}{=}\frac{1}{2\pi\sigma_{x}\sigma_{y}}\exp\left[  -\frac
{1}{2\sigma_{x}^{2}}\left(  x-\mu_{x}\right)  ^{2}-\frac{1}{2\sigma_{y}^{2}%
}y^{2}\right]  \text{,} \label{PGAUSS}%
\end{equation}
with $\sigma_{x}$ and $\sigma_{y}$ in $%
%TCIMACRO{\U{211d} }%
%BeginExpansion
\mathbb{R}
%EndExpansion
_{0}^{+}$ and $\mu_{x}$ in $%
%TCIMACRO{\U{211d} }%
%BeginExpansion
\mathbb{R}
%EndExpansion
$. The Gaussian here is two dimensional in its microscopic space $(x$, $y)$
but three dimensional in its macroscopic (contextual or conditionally given
parameters) space. We will now relax this conditionality to explore the space
of Gaussians described by $\mu_{x}$, $\sigma_{x}$\ and $\sigma_{y}$. The
infinitesimal Fisher-Rao line element $ds_{3D}^{2}$ reads,%
\begin{equation}
ds_{3D}^{2}=\sum_{l\text{, }m=1}^{3}g_{lm}^{\left(  3D\right)  }\left(
\theta\right)  d\theta^{l}d\theta^{m}=\frac{1}{\sigma_{x}^{2}}d\mu_{x}%
^{2}+\frac{2}{\sigma_{x}^{2}}d\sigma_{x}^{2}+\frac{2}{\sigma_{y}^{2}}%
d\sigma_{y}^{2}\text{,}%
\end{equation}
where the Fisher-Rao information metric $g_{lm}\left(  \theta\right)  $ is
defined as \cite{amari},%
\begin{equation}
g_{lm}^{\left(  3D\right)  }\left(  \theta\right)  \overset{\text{def}}{=}\int
dxdyp\left(  x\text{, }y|\mu_{x}\text{, }\sigma_{x}\text{, }\sigma_{y}\right)
\frac{\partial\log p\left(  x\text{, }y|\mu_{x}\text{, }\sigma_{x}\text{,
}\sigma_{y}\right)  }{\partial\theta^{l}}\frac{\partial\log p\left(  x\text{,
}y|\mu_{x}\text{, }\sigma_{x}\text{, }\sigma_{y}\right)  }{\partial\theta^{l}%
}\text{,}%
\end{equation}
with $\theta\equiv\left(  \theta^{1}\text{, }\theta^{2}\text{, }\theta
^{3}\right)  \overset{\text{def}}{=}\left(  \mu_{x}\text{, }\sigma_{x}\text{,
}\sigma_{y}\right)  $ and where $g_{lm}^{\left(  3D\right)  }\left(
\theta\right)  $ has the following matrix representation,%
\begin{equation}
g_{lm}^{\left(  3D\right)  }\left(  \theta\right)  =\left(
\begin{array}
[c]{ccc}%
\frac{1}{\sigma_{x}^{2}} & 0 & 0\\
0 & \frac{2}{\sigma_{x}^{2}} & 0\\
0 & 0 & \frac{2}{\sigma_{y}^{2}}%
\end{array}
\right)  \text{.} \label{m3}%
\end{equation}
Using (\ref{m3}), it follows that the non-vanishing Christoffel connection
coefficients $\Gamma_{ij}^{k}$ \cite{landau},%
\begin{equation}
\Gamma_{ij}^{k}=\frac{1}{2}g^{km}\left(  \partial_{i}g_{mj}+\partial_{j}%
g_{im}-\partial_{m}g_{ij}\right)  \text{,} \label{symbols}%
\end{equation}
are given by,%
\begin{equation}
\Gamma_{12}^{1}=\Gamma_{21}^{1}=-\frac{1}{\sigma_{x}}\text{, }\Gamma_{11}%
^{2}=\frac{1}{2\sigma_{x}}\text{, }\Gamma_{22}^{2}=-\frac{1}{\sigma_{x}%
}\text{, }\Gamma_{33}^{3}=-\frac{1}{\sigma_{y}}\text{.} \label{crist}%
\end{equation}
The scalar curvature $\mathcal{R}^{\left(  3D\right)  }$ of the statistical
manifold of probability distributions in (\ref{PGAUSS}) is given by,%
\begin{equation}
\mathcal{R}^{\left(  3D\right)  }=g^{11}\left(  \theta\right)  R_{11}%
+g^{22}\left(  \theta\right)  R_{22}+g^{33}\left(  \theta\right)
R_{33}\text{,} \label{riccis}%
\end{equation}
with $g^{lm}g_{mk}=\delta_{k}^{l}$ and the Ricci curvature tensor $R_{ij}$
defined as \cite{landau},%
\begin{equation}
R_{ij}\overset{\text{def}}{=}\partial_{k}\Gamma_{ij}^{k}-\partial_{j}%
\Gamma_{ik}^{k}+\Gamma_{ij}^{k}\Gamma_{kn}^{n}-\Gamma_{ik}^{m}\Gamma_{jm}%
^{k}\text{.} \label{ricci}%
\end{equation}
Substituting (\ref{crist}) into (\ref{ricci}), it turns out that the non
vanishing components of $R_{ij}$ are,%
\begin{equation}
R_{11}=-\frac{1}{2\sigma_{x}^{2}}\text{, }R_{22}=-\frac{1}{\sigma_{x}^{2}%
}\text{.}%
\end{equation}
Thus, the scalar curvature becomes $\mathcal{R}^{\left(  3D\right)  }=-1$. We
finally point out the only non-vanishing component of the Riemann curvature
tensor $R_{\mu\nu\rho}^{\alpha}$ \cite{landau},%
\begin{equation}
R_{\mu\nu\rho}^{\alpha}\overset{\text{def}}{=}\partial_{\nu}\Gamma_{\mu\rho
}^{\alpha}-\partial_{\rho}\Gamma_{\mu\nu}^{\alpha}+\Gamma_{\beta\nu}^{\alpha
}\Gamma_{\mu\rho}^{\beta}-\Gamma_{\beta\rho}^{\alpha}\Gamma_{\mu\nu}^{\beta
}\text{,}%
\end{equation}
is given by,%
\begin{equation}
R_{212}^{1}=-\frac{1}{\sigma_{x}^{2}}\text{.}%
\end{equation}
As a final remark, note that
\begin{equation}
\mathcal{R}^{\left(  3D\right)  }\equiv\left(  R_{1212}+R_{2121}\right)
g^{11}\left(  \theta\right)  g^{22}\left(  \theta\right)  =-1\text{,}
\label{ricci1}%
\end{equation}
in agreement with Eq. (\ref{riccis}).

\subsection{The 2D Statistical Model}

The probability distributions $p\left(  x\text{, }y\text{; }\mu_{x}\text{,
}\sigma\right)  $ that characterize the $2D$ Gaussian statistical model are
given by,%
\begin{equation}
p\left(  x\text{, }y\text{; }\mu_{x}\text{, }\sigma\right)  \overset
{\text{def}}{=}\frac{1}{2\pi\Sigma^{2}}\exp\left[  -\frac{1}{2\sigma^{2}%
}\left(  x-\mu_{x}\right)  ^{2}-\frac{\sigma^{2}}{2\Sigma^{4}}y^{2}\right]
\text{,} \label{GP}%
\end{equation}
with $\sigma$ in $%
%TCIMACRO{\U{211d} }%
%BeginExpansion
\mathbb{R}
%EndExpansion
_{0}^{+}$ and $\mu_{x}$ in $%
%TCIMACRO{\U{211d} }%
%BeginExpansion
\mathbb{R}
%EndExpansion
$. The probability distribution $p\left(  x\text{, }y\text{; }\mu_{x}\text{,
}\sigma\right)  $ may be obtained from $p\left(  x\text{, }y|\mu_{x}\text{,
}\sigma_{x}\text{, }\sigma_{y}\right)  $ with the addition of the following
\emph{macroscopic constraint},%
\begin{equation}
\sigma_{x}\sigma_{y}=\Sigma^{2}\text{, } \label{MC}%
\end{equation}
where $\Sigma^{2}$ is a constant belonging to $%
%TCIMACRO{\U{211d} }%
%BeginExpansion
\mathbb{R}
%EndExpansion
_{0}^{+}$ and $\sigma_{x}\equiv\sigma$. The macroscopic constraint (\ref{MC})
resembles the quantum mechanical canonical minimum uncertainty relation where
$x$ denotes the position of a particle and $y$ its conjugate momentum. The
infinitesimal Fisher-Rao line element $ds_{2D}^{2}$ reads,%
\begin{equation}
ds_{2D}^{2}=\sum_{l\text{, }m=1}^{2}g_{lm}^{\left(  2D\right)  }\left(
\theta\right)  d\theta^{l}d\theta^{m}=\frac{1}{\sigma^{2}}d\mu_{x}^{2}%
+\frac{4}{\sigma^{2}}d\sigma^{2}\text{,}%
\end{equation}
where the Fisher-Rao information metric $g_{lm}\left(  \theta\right)  $ is
defined as,%
\begin{equation}
g_{lm}^{\left(  2D\right)  }\left(  \theta\right)  \overset{\text{def}}{=}\int
dxdyp\left(  x\text{, }y\text{; }\mu_{x}\text{, }\sigma\right)  \frac
{\partial\log p\left(  x\text{, }y\text{; }\mu_{x}\text{, }\sigma\right)
}{\partial\theta^{l}}\frac{\partial\log p\left(  x\text{, }y\text{; }\mu
_{x}\text{, }\sigma\right)  }{\partial\theta^{l}}\text{,}%
\end{equation}
with $\theta\equiv\left(  \theta^{1}\text{, }\theta^{2}\right)  \overset
{\text{def}}{=}\left(  \mu_{x}\text{, }\sigma\right)  $ and where
$g_{lm}^{\left(  2D\right)  }\left(  \theta\right)  $ has the following matrix
representation,%
\begin{equation}
g_{lm}^{\left(  2D\right)  }\left(  \theta\right)  =\frac{1}{\sigma^{2}%
}\left(
\begin{array}
[c]{cc}%
1 & 0\\
0 & 4
\end{array}
\right)  \text{.} \label{m4}%
\end{equation}
Using (\ref{m4}), it follows that the non-vanishing Christoffel connection
coefficients $\Gamma_{ij}^{k}$ are given by,%
\begin{equation}
\Gamma_{12}^{1}=\Gamma_{21}^{1}=-\frac{1}{\sigma}\text{, }\Gamma_{11}%
^{2}=\frac{1}{4\sigma}\text{, }\Gamma_{22}^{2}=-\frac{1}{\sigma}\text{.}
\label{crist2}%
\end{equation}
The scalar curvature $\mathcal{R}^{\left(  2D\right)  }$ of the probability
distributions in (\ref{GP}) is given by,%
\begin{equation}
\mathcal{R}^{\left(  2D\right)  }=g^{11}\left(  \theta\right)  R_{11}%
+g^{22}\left(  \theta\right)  R_{22}=-\frac{1}{2}\text{,} \label{riccis2}%
\end{equation}
with $g^{lm}g_{mk}=\delta_{k}^{l}$ and where the only non-vanishing Ricci
curvature tensor component $R_{ij}$ is,%
\begin{equation}
R_{11}=-\frac{1}{4\sigma^{2}}\text{, }R_{22}=-\frac{1}{\sigma^{2}}\text{.}%
\end{equation}
Observe that the only non-vanishing component of the Riemann curvature tensor
$R_{\mu\nu\rho}^{\alpha}$ is,%
\begin{equation}
R_{212}^{1}=-\frac{1}{\sigma^{2}}\text{,}%
\end{equation}
thus,
\begin{equation}
\mathcal{R}^{\left(  2D\right)  }\equiv\left(  R_{1212}+R_{2121}\right)
g^{11}\left(  \theta\right)  g^{22}\left(  \theta\right)  =-\frac{1}%
{2}\text{,} \label{ricci2}%
\end{equation}
in agreement with Eq. (\ref{riccis2}). From (\ref{ricci1}) and (\ref{ricci2}),
it turns out that the $3D$ statistical model is \emph{globally} more
negatively curved than the $2D$ statistical model.

\section{Geodesic Motion on Curved Statistical Manifolds}

In this Section, we present the geodesic paths on the curved statistical
manifolds underlying the entropic dynamics of the three and two-dimensional
Gaussian statistical models. Such paths are obtained by integrating the
geodesic equations given by \cite{landau},%
\begin{equation}
\frac{d^{2}\theta^{k}\left(  s\right)  }{d\tau^{2}}+\Gamma_{lm}^{k}\left(
\theta\right)  \frac{d\theta^{l}}{d\tau}\frac{d\theta^{m}}{d\tau}=0\text{,}
\label{GEE}%
\end{equation}
where $\Gamma_{lm}^{k}\left(  \theta\right)  $ are the Christoffel symbols
defined in (\ref{symbols}).

\subsection{The Three-dimensional Case}

Substituting (\ref{crist}) into (\ref{GEE}), the set of nonlinear and coupled
ordinary differential equations in (\ref{GEE}) reads,%
\begin{align}
0  &  =\frac{d^{2}\mu_{x}}{d\tau^{2}}-\frac{2}{\sigma_{x}}\frac{d\mu_{x}%
}{d\tau}\frac{d\sigma_{x}}{d\tau}\text{,}\nonumber\\
& \nonumber\\
0  &  =\frac{d^{2}\sigma_{x}}{d\tau^{2}}+\frac{1}{2\sigma_{x}}\left(
\frac{d\mu_{x}}{d\tau}\right)  ^{2}-\frac{1}{\sigma_{x}}\left(  \frac{d\sigma
}{d\tau}\right)  ^{2}\text{,}\nonumber\\
& \nonumber\\
0  &  =\frac{d^{2}\sigma_{y}}{d\tau^{2}}-\frac{1}{\sigma_{y}}\left(
\frac{d\sigma_{y}}{d\tau}\right)  ^{2}\text{.}%
\end{align}
A suitable family of geodesic paths fulfilling the geodesic equations above is
given by (technical details appear in Appendix A),%
\begin{equation}
\mu_{x}\left(  \tau\right)  =\frac{\left(  \mu_{0}+2\sigma_{0}\right)  \left[
1+\exp\left(  2\sigma_{0}\lambda_{+}^{\prime}\tau\right)  \right]
-4\sigma_{0}}{1+\exp\left(  2\sigma_{0}\lambda_{+}^{\prime}\tau\right)
}\text{, }\sigma_{x}\left(  \tau\right)  =\frac{2\sigma_{0}\exp\left(
\sigma_{0}\lambda_{+}^{\prime}\tau\right)  }{1+\exp\left(  2\sigma_{0}%
\lambda_{+}^{\prime}\tau\right)  }\text{,} \label{cac1}%
\end{equation}
and,%
\begin{equation}
\sigma_{y}\left(  \tau\right)  =\sigma_{0}^{\prime}\exp\left(  -\lambda
_{f}\tau\right)  \text{,} \label{cac2}%
\end{equation}
where $\mu_{0}\overset{\text{def}}{=}\mu_{x}\left(  0\right)  $, $\sigma
_{0}\overset{\text{def}}{=}\sigma_{x}\left(  0\right)  $, $\sigma_{0}^{\prime
}\overset{\text{def}}{=}\sigma_{y}\left(  0\right)  $, $\varepsilon
\overset{\text{def}}{=}\sigma_{y}\left(  \tau_{f}\right)  $, $\lambda
_{f}\overset{\text{def}}{=}\frac{1}{\tau_{f}}\log\left(  \frac{\sigma_{0}%
}{\varepsilon}\right)  $ and $\lambda_{+}^{\prime}\in%
%TCIMACRO{\U{211d} }%
%BeginExpansion
\mathbb{R}
%EndExpansion
^{+}$ (see Appendix A).

\subsection{The Two-Dimensional Case}

Substituting (\ref{crist2}) into (\ref{GEE}), the set of nonlinear and coupled
ordinary differential equations in (\ref{GEE}) reads,%
\begin{align}
0  &  =\frac{d^{2}\mu_{x}}{d\tau^{2}}-\frac{2}{\sigma}\frac{d\mu_{x}}{d\tau
}\frac{d\sigma}{d\tau}\text{,}\nonumber\\
& \nonumber\\
0  &  =\frac{d^{2}\sigma}{d\tau^{2}}+\frac{1}{4\sigma}\left(  \frac{d\mu_{x}%
}{d\tau}\right)  ^{2}-\frac{1}{\sigma}\left(  \frac{d\sigma}{d\tau}\right)
^{2}\text{.}%
\end{align}
A suitable family of geodesic paths fulfilling the geodesic equations above is
given by,%
\begin{equation}
\mu_{x}\left(  \tau\right)  =\frac{\left(  \mu_{0}+2\sigma_{0}\right)  \left[
1+\exp\left(  2\sigma_{0}\lambda_{+}\tau\right)  \right]  -4\sigma_{0}}%
{1+\exp\left(  2\sigma_{0}\lambda_{+}\tau\right)  }\text{,} \label{cac3}%
\end{equation}
and,%
\begin{equation}
\sigma\left(  \tau\right)  =\frac{2\sigma_{0}\exp\left(  \sigma_{0}\lambda
_{+}\tau\right)  }{1+\exp\left(  2\sigma_{0}\lambda_{+}\tau\right)  }\text{,}
\label{cac4}%
\end{equation}
where $\mu_{0}\overset{\text{def}}{=}\mu_{x}\left(  0\right)  $, $\sigma
_{0}\overset{\text{def}}{=}\sigma\left(  0\right)  $ and $\lambda_{+}=$
$\frac{\lambda_{+}^{\prime}}{\sqrt{2}}$ belongs to $%
%TCIMACRO{\U{211d} }%
%BeginExpansion
\mathbb{R}
%EndExpansion
^{+}$ (see Appendix A).

\section{Indicators of Chaoticity}

In this Section, the chaotic properties of the information-constrained
(entropic) dynamics on the underlying curved statistical manifolds are
quantified by means of the Information Geometric Entropy and the Jacobi vector
field intensity. The relevance of such quantities as suitable indicators of
chaoticity was also investigated in \cite{carloIJTP}.

\subsection{Information Geometric Entropy}

We point out that a suitable indicator of temporal complexity within the IGAC
framework is provided by the IGE $\mathcal{S}_{\mathcal{M}_{s}}\left(
\tau\right)  $ \cite{AMC},%
\begin{equation}
\mathcal{S}_{\mathcal{M}_{s}}\left(  \tau\right)  \overset{\text{def}}{=}%
\log\widetilde{\emph{vol}}\left[  \mathcal{D}_{\theta}^{\left(
\text{geodesic}\right)  }\left(  \tau\right)  \right]  \text{.}%
\end{equation}
The average dynamical statistical volume $\widetilde{\emph{vol}}\left[
\mathcal{D}_{\Theta}^{\left(  \text{geodesic}\right)  }\left(  \tau\right)
\right]  $ is defined as,%
\begin{equation}
\widetilde{\emph{vol}}\left[  \mathcal{D}_{\theta}^{\left(  \text{geodesic}%
\right)  }\left(  \tau\right)  \right]  \overset{\text{def}}{=}\lim
_{\tau\rightarrow\infty}\left(  \frac{1}{\tau}\int_{0}^{\tau}d\tau^{\prime
}\emph{vol}\left[  \mathcal{D}_{\theta}^{\left(  \text{geodesic}\right)
}\left(  \tau^{\prime}\right)  \right]  \right)  \text{,} \label{rhs}%
\end{equation}
where the "tilde" symbol denotes the operation of temporal average. For the
sake of clarity, we point out that in the RHS of (\ref{rhs}), we intend to
preserve the temporal-dependence by considering the asymptotic leading term in
the limit of $\tau$ approaching infinity. For a $n$-dimensional manifold, the
volume $\emph{vol}\left[  \mathcal{D}_{\theta}^{\left(  \text{geodesic}%
\right)  }\left(  \tau^{\prime}\right)  \right]  $ in (\ref{rhs}) is given by,%
\begin{equation}
vol\left[  \mathcal{D}_{\theta}^{\left(  \text{geodesic}\right)  }\left(
\tau^{\prime}\right)  \right]  \overset{\text{def}}{=}\int_{\mathcal{D}%
_{\theta}^{\left(  \text{geodesic}\right)  }\left(  \tau^{\prime}\right)
}\rho_{\left(  \mathcal{M}_{s}\text{, }g\right)  }\left(  \theta
^{1}\text{,..., }\theta^{n}\right)  d^{n}\theta\text{,} \label{v}%
\end{equation}
where $\rho_{\left(  \mathcal{M}_{s}\text{, }g\right)  }\left(  \theta
^{1}\text{,..., }\theta^{n}\right)  $ is the so-called Fisher density and
equals the square root of the determinant of the metric tensor $g_{\mu\nu
}\left(  \theta\right)  $,%
\begin{equation}
\rho_{\left(  \mathcal{M}_{s}\text{, }g\right)  }\left(  \theta^{1}\text{,...,
}\theta^{n}\right)  \overset{\text{def}}{=}\sqrt{g\left(  \left(  \theta
^{1}\text{,..., }\theta^{n}\right)  \right)  }\text{.}%
\end{equation}
The integration space $\mathcal{D}_{\theta}^{\left(  \text{geodesic}\right)
}\left(  \tau^{\prime}\right)  $ in (\ref{v}) is defined as follows,%
\begin{equation}
\mathcal{D}_{\theta}^{\left(  \text{geodesic}\right)  }\left(  \tau^{\prime
}\right)  \overset{\text{def}}{=}\left\{  \theta\equiv\left(  \theta
^{1}\text{,..., }\theta^{n}\right)  :\theta^{k}\left(  0\right)  \leq
\theta^{k}\leq\theta^{k}\left(  \tau^{\prime}\right)  \right\}  \text{,}
\label{is}%
\end{equation}
where $k=1$,.., $n$ and $\theta^{k}\equiv\theta^{k}\left(  s\right)  $ with
$0\leq s\leq\tau^{\prime}$ such that,%
\begin{equation}
\frac{d^{2}\theta^{k}\left(  s\right)  }{ds^{2}}+\Gamma_{lm}^{k}\frac
{d\theta^{l}}{ds}\frac{d\theta^{m}}{ds}=0\text{.}%
\end{equation}
The integration space $\mathcal{D}_{\theta}^{\left(  \text{geodesic}\right)
}\left(  \tau^{\prime}\right)  $ in (\ref{is}) is a $n$-dimensional subspace
of the whole (permitted) parameter space $\mathcal{D}_{\theta}^{\left(
\text{tot}\right)  }$. The elements of $\mathcal{D}_{\theta}^{\left(
\text{geodesic}\right)  }\left(  \tau^{\prime}\right)  $ are the
$n$-dimensional macrovariables $\left\{  \theta\right\}  $ whose components
$\theta^{k}$ are bounded by specified limits of integration $\theta^{k}\left(
0\right)  $ and $\theta^{k}\left(  \tau^{\prime}\right)  $ with $k=1$,.., $n$.
The limits of integration are obtained via integration of the $n$-dimensional
set of coupled nonlinear second order ordinary differential equations
characterizing the geodesic equations. Formally, the IGE $\mathcal{S}%
_{\mathcal{M}_{s}}\left(  \tau\right)  $ is defined in terms of a averaged
parametric $\left(  n+1\right)  $-fold integral ($\tau$ is the parameter) over
the multidimensional geodesic paths connecting $\theta\left(  0\right)  $ to
$\theta\left(  \tau\right)  $.

In the cases being investigated, using (\ref{cac1}) and (\ref{cac2}) it turns
out that the $\mathcal{S}_{\mathcal{M}_{s}}^{\left(  3D\text{ Model}\right)
}$ of the $3D$ statistical model reads,%
\begin{equation}
\mathcal{S}_{\mathcal{M}_{s}}^{\left(  3D\text{ Model}\right)  }%
=\log\mathcal{V}_{\mathcal{M}_{s}}^{\left(  3D\text{ Model}\right)  }\left(
\tau\right)  \text{,}%
\end{equation}
with,%
\begin{equation}
\mathcal{V}_{\mathcal{M}_{s}}^{\left(  3D\text{ Model}\right)  }\left(
\tau\right)  =\frac{1}{\sigma_{0}^{3}\lambda_{+}^{\prime}{}^{2}}\frac
{\exp\left(  \sigma_{0}\lambda_{+}^{\prime}\tau\right)  }{\tau}\left[
\begin{array}
[c]{c}%
\left(  2\sigma_{0}+\mu_{0}\right)  \sigma_{0}\lambda_{f}\lambda_{+}^{\prime
}\tau+\left(  2\sigma_{0}-\mu_{0}\right)  \sigma_{0}\lambda_{f}\lambda
_{+}^{\prime}\tau e^{-2\sigma_{0}\lambda^{\prime}\tau}+\\
\\
-\left(  \lambda_{f}+\lambda_{+}^{\prime}\sigma_{0}\ln\sigma_{0}^{\prime
}\right)  \left(  2\sigma_{0}+\mu_{0}\right)  +\\
\\
-\left(  \sigma_{0}\lambda_{+}^{\prime}\ln\sigma_{0}^{\prime}-\lambda
_{f}\right)  \left(  2\sigma_{0}-\mu_{0}\right)  e^{-2\sigma_{0}\tau
\lambda^{\prime}}%
\end{array}
\right]  \text{.}%
\end{equation}
In the asymptotic limit, we get%
\begin{equation}
\mathcal{V}_{\mathcal{M}_{s}}^{\left(  3D\text{ Model}\right)  }\left(
\tau\right)  \overset{\tau\gg1}{\approx}\left[  \left(  \frac{\lambda_{f}%
}{\lambda_{+}^{\prime}}\cdot\frac{\mu_{0}+2\sigma_{0}}{\sigma_{0}^{2}}\right)
\exp\left(  \sigma_{0}\lambda_{+}^{\prime}\tau\right)  \right]  \text{,}%
\end{equation}
that is,%
\begin{equation}
\mathcal{S}_{\mathcal{M}_{s}}^{\left(  3D\text{ Model}\right)  }\left(
\tau\right)  \overset{\tau\gg1}{\approx}\lambda_{+}^{\prime}\tau\text{.}
\label{IGE2}%
\end{equation}
Similarly, using (\ref{cac3}) and (\ref{cac4}), it follows that the
$\mathcal{S}_{\mathcal{M}_{s}}^{\left(  2D\text{ Model}\right)  }$ of the $2D$
model becomes,%
\begin{equation}
\mathcal{S}_{\mathcal{M}_{s}}^{\left(  2D\text{ Model}\right)  }%
=\log\mathcal{V}_{\mathcal{M}_{s}}^{\left(  2D\text{ Model}\right)  }\left(
\tau\right)  \text{,}%
\end{equation}
with,%
\begin{equation}
\mathcal{V}_{\mathcal{M}_{s}}^{\left(  2D\text{ Model}\right)  }\left(
\tau\right)  =\frac{1}{\lambda_{+}\sigma_{0}^{2}}\frac{\left(  \mu_{0}%
+2\sigma_{0}\right)  +\left(  2\sigma_{0}-\mu_{0}\right)  \exp\left(
-2\sigma_{0}\lambda_{+}\tau\right)  }{\tau\exp\left(  -\sigma_{0}\lambda
_{+}\tau\right)  }\text{.}%
\end{equation}
In the asymptotic limit, we obtain%
\begin{equation}
\mathcal{V}_{\mathcal{M}_{s}}^{\left(  2D\text{ Model}\right)  }\left(
\tau\right)  \overset{\tau\gg1}{\approx}\left[  \left(  \frac{\mu_{0}%
+2\sigma_{0}}{\sigma_{0}^{2}\lambda_{+}}\right)  \frac{\exp\left(  \sigma
_{0}\lambda_{+}\tau\right)  }{\tau}\right]  \text{,}%
\end{equation}
and $\mathcal{S}_{\mathcal{M}_{s}}^{\left(  2D\text{ Model}\right)  }$
becomes,%
\begin{equation}
\mathcal{S}_{\mathcal{M}_{s}}^{\left(  2D\text{ Model}\right)  }\left(
\tau\right)  \overset{\tau\gg1}{\approx}\lambda_{+}\tau\text{.} \label{IGE1}%
\end{equation}
Combining (\ref{IGE1}) and (\ref{IGE2}), it finally turns out that%
\begin{equation}
\mathcal{S}_{\mathcal{M}_{s}}^{\left(  2D\text{ Model}\right)  }\left(
\tau\right)  \overset{\tau\gg1}{\approx}\left[  \left(  \frac{\lambda_{+}%
}{\lambda_{+}^{\prime}}\right)  \cdot\mathcal{S}_{\mathcal{M}_{s}}^{\left(
3D\text{ Model}\right)  }\left(  \tau\right)  \right]  \text{ with }%
\frac{\lambda_{+}}{\lambda_{+}^{\prime}}=\frac{1}{\sqrt{2}}<1\text{.}
\label{IGEfinal}%
\end{equation}
Eq. (\ref{IGEfinal}) is quite interesting since it quantitatively shows that
the IGE\ is softened when approaching the two-dimensional case from the
three-dimensional case via the introduction of the macroscopic constraint
(\ref{MC}) that is reminiscent of Heisenberg's minimum uncertainty relation
where\textbf{\ }$x$\textbf{\ }denotes the position of a particle and $y$\ its
conjugate momentum.

\subsection{Jacobi Vector Field Intensity}

The Jacobi-Levi-Civita (JLC) equation of geodesic deviation is a complicated
second-order system of linear ordinary differential equations. It describes
the geodesic spread on curved manifolds of a pair of nearby freely falling
particles travelling on trajectories $\theta^{\rho}\left(  \tau\right)  $ and
$\theta^{\prime\rho}\left(  \tau\right)  \overset{\text{def}}{=}\theta^{\rho
}\left(  \tau\right)  +\delta\theta^{\rho}\left(  \tau\right)  $. The JLC
equation is given by \cite{felice},%
\begin{equation}
\frac{D^{2}J^{k}}{D\tau^{2}}+R_{nml}^{k}\frac{\partial\theta^{n}}{\partial
\tau}J^{m}\frac{\partial\theta^{l}}{\partial\tau}=0\text{, } \label{jlc}%
\end{equation}
with $k=1$, $2$, $3$ (in the three-dimensional case) and where the covariant
derivative $\frac{D\theta^{\mu}\left(  \tau\right)  }{D\tau}$ along the curve
$\theta^{\mu}\left(  \tau\right)  $ is defined as,%
\begin{equation}
\frac{D\theta^{\mu}\left(  \tau\right)  }{D\tau}\overset{\text{def}}{=}%
\frac{d\Theta^{\mu}\left(  \tau\right)  }{d\tau}+\Gamma_{\nu\rho}^{\mu}%
\frac{d\Theta^{\rho}}{d\tau}\Theta^{\nu}\text{.}%
\end{equation}
The Jacobi vector field components $J^{k}$ are given by,%
\begin{equation}
J^{k}\equiv\delta_{\lambda_{k}}\theta^{k}\overset{\text{def}}{=}\left(
\frac{\partial\theta^{k}\left(  \tau\text{; }\lambda_{k}\right)  }%
{\partial\lambda_{k}}\right)  _{\tau}\delta\lambda_{k}\text{,} \label{jacobi}%
\end{equation}
and $R_{\alpha\beta\gamma\delta}$ is the Riemann curvature tensor defined as
\cite{felice},%
\begin{equation}
R_{\mu\nu\rho}^{\alpha}\overset{\text{def}}{=}\partial_{\nu}\Gamma_{\mu\rho
}^{\alpha}-\partial_{\rho}\Gamma_{\mu\nu}^{\alpha}+\Gamma_{\beta\nu}^{\alpha
}\Gamma_{\mu\rho}^{\beta}-\Gamma_{\beta\rho}^{\alpha}\Gamma_{\mu\nu}^{\beta
}\text{.}%
\end{equation}
The covariant derivative $\frac{D^{2}J^{\mu}}{D\tau^{2}}$ in (\ref{jlc}) is
defined as \cite{ohanian},
\begin{equation}
\frac{D^{2}J^{\mu}}{D\tau^{2}}=\frac{d^{2}J^{\mu}}{d\tau^{2}}+2\Gamma
_{\alpha\beta}^{\mu}\frac{dJ^{\alpha}}{d\tau}\frac{d\Theta^{\beta}}{d\tau
}+\Gamma_{\alpha\beta}^{\mu}J^{\alpha}\frac{d^{2}\Theta^{\beta}}{d\tau^{2}%
}+\Gamma_{\alpha\beta\text{, }\nu}^{\mu}\frac{d\Theta^{\nu}}{d\tau}%
\frac{d\Theta^{\beta}}{d\tau}J^{\alpha}+\Gamma_{\alpha\beta}^{\mu}\Gamma
_{\rho\sigma}^{\alpha}\frac{d\Theta^{\sigma}}{d\tau}\frac{d\Theta^{\beta}%
}{d\tau}J^{\rho}\text{.} \label{1d}%
\end{equation}
Equation (\ref{jlc}) forms a system of coupled ordinary differential equations
\textit{linear} in the components of the deviation vector field (\ref{jacobi})
but\textit{\ nonlinear} in derivatives of the metric tensor $g_{ij}\left(
\theta\right)  $. It describes the linearized geodesic flow: the linearization
ignores the relative velocity of the geodesics. When the geodesics are
neighboring but their relative velocity is arbitrary, the corresponding
geodesic deviation equation is the so-called generalized Jacobi equation
\cite{chicone}. The nonlinearity is due to the existence of velocity-dependent
terms in the system. Neighboring geodesics accelerate relative to each other
with a rate directly measured by the curvature tensor $R_{\alpha\beta
\gamma\delta}$.

Omitting technical details (see Appendix B) and setting $\Lambda_{3D}%
\overset{\text{def}}{=}\lambda_{3D}\sigma_{0}$ with $\lambda_{3D}=\lambda
_{+}^{\prime}$, the JLC equation for $J^{1}$ becomes,%
\begin{equation}
\frac{d^{2}J^{1}}{d\tau^{2}}+2\Lambda_{3D}\frac{dJ^{1}}{d\tau}-8\Lambda
_{3D}\exp\left(  -\Lambda_{3D}\tau\right)  \frac{dJ^{2}}{d\tau}-8\Lambda
_{3D}\exp\left(  -2\Lambda_{3D}\tau\right)  J^{1}-8\Lambda_{3D}^{2}\exp\left(
-\Lambda_{3D}\tau\right)  J^{2}=0\text{.}%
\end{equation}
Similarly, it can be shown that the JLC equations for $J^{2}$ and $J^{3}$
read,%
\begin{equation}
\frac{d^{2}J^{2}}{d\tau^{2}}+4\Lambda_{3D}\exp\left(  -\Lambda_{3D}%
\tau\right)  \frac{dJ^{1}}{d\tau}+2\Lambda_{3D}\frac{dJ^{2}}{d\tau}%
+\Lambda_{3D}^{2}J^{2}=0\text{,}%
\end{equation}
and,%
\begin{equation}
\frac{d^{2}J^{3}}{d\tau^{2}}+2\lambda_{f}\frac{dJ^{3}}{d\tau}+\lambda_{f}%
^{2}J^{3}=0\text{,}%
\end{equation}
respectively. In the asymptotic limit, it can be shown that the three
equations to integrate become (see Appendix B)%
\begin{align}
\frac{d^{2}J^{1}}{d\tau^{2}}+2\Lambda_{3D}\frac{dJ^{1}}{d\tau}  &  =0\text{,
}\nonumber\\
& \nonumber\\
\frac{d^{2}J^{2}}{d\tau^{2}}+2\Lambda_{3D}\frac{dJ^{2}}{d\tau}+\Lambda
_{3D}^{2}J^{2}  &  =0\text{,}\nonumber\\
& \nonumber\\
\frac{d^{2}J^{3}}{d\tau^{2}}+2\lambda_{f}\frac{dJ^{3}}{d\tau}+\lambda_{f}%
^{2}J^{3}  &  =0\text{.}%
\end{align}
The asymptotic solutions are given by,%
\begin{align}
J^{1}\left(  \tau\right)   &  =C_{1}^{\left(  1\right)  }+C_{2}^{\left(
1\right)  }\exp\left(  -2\Lambda_{3D}\tau\right)  \text{, }\nonumber\\
& \nonumber\\
J^{2}\left(  \tau\right)   &  =C_{1}^{\left(  2\right)  }\exp\left(
-\Lambda_{3D}\tau\right)  +C_{2}^{\left(  2\right)  }\tau\exp\left(
-\Lambda_{3D}\tau\right)  \text{, }\nonumber\\
& \nonumber\\
J^{3}\left(  \tau\right)   &  =C_{1}^{\left(  3\right)  }\exp\left(
-\lambda_{f}\tau\right)  +C_{2}^{\left(  3\right)  }\tau\exp\left(
-\lambda_{f}\tau\right)  \text{,} \label{j1j}%
\end{align}
where $C_{k^{\prime}}^{\left(  k\right)  }$ with $k=1$, $2$, $3$ and
$k^{\prime}=1$, $2$ are \emph{real} integration constants. The Jacobi vector
field intensity $J_{\mathcal{M}_{s}}^{\left(  3D\right)  }\left(  \tau\right)
$ is defined as,%
\begin{equation}
J_{\mathcal{M}_{s}}^{\left(  3D\right)  }\left(  \tau\right)  \overset
{\text{def}}{=}\left[  \frac{\left[  J^{1}\left(  \tau\right)  \right]  ^{2}%
}{\sigma_{x}^{2}\left(  \tau\right)  }+\frac{2\left[  J^{2}\left(
\tau\right)  \right]  ^{2}}{\sigma_{x}^{2}\left(  \tau\right)  }%
+\frac{2\left[  J^{3}\left(  \tau\right)  \right]  ^{2}}{\sigma_{y}^{2}\left(
\tau\right)  }\right]  ^{\frac{1}{2}}\text{.}%
\end{equation}
Eqs. (\ref{cac1}), (\ref{cac2}) and (\ref{j1j}) imply that,%
\begin{equation}
\frac{\left[  J^{1}\left(  \tau\right)  \right]  ^{2}}{\sigma_{x}^{2}\left(
\tau\right)  }\approx\frac{\left[  C_{1}^{\left(  1\right)  }\right]  ^{2}%
}{4\sigma_{0}^{2}}\exp\left(  2\Lambda_{3D}\tau\right)  \text{, }%
\frac{2\left[  J^{2}\left(  \tau\right)  \right]  ^{2}}{\sigma_{x}^{2}\left(
\tau\right)  }\approx\frac{\left[  C_{2}^{\left(  2\right)  }\right]  ^{2}%
}{2\sigma_{0}^{2}}\tau^{2}\text{ and, }\frac{2\left[  J^{3}\left(
\tau\right)  \right]  ^{2}}{\sigma_{y}^{2}\left(  \tau\right)  }\approx
\frac{2\left[  C_{2}^{\left(  3\right)  }\right]  ^{2}}{\sigma_{0}^{\prime2}%
}\tau^{2}\text{.}%
\end{equation}
It finally follows that $J_{\mathcal{M}_{s}}^{\left(  3D\right)  }\left(
\tau\right)  $ reads,%
\begin{equation}
J_{\mathcal{M}_{s}}^{\left(  3D\right)  }\left(  \tau\right)  \approx
\frac{C_{1}^{\left(  1\right)  }}{2\sigma_{0}}\exp\left(  \Lambda_{3D}%
\tau\right)  \text{.}%
\end{equation}
Consider now the two-dimensional statistical model. Omitting technical details
(see Appendix B) and setting $\Lambda_{2D}\overset{\text{def}}{=}\lambda
_{2D}\sigma_{0}$ with $\lambda_{2D}=\lambda_{+}$, the two JLC equations for
$J^{1}$ and $J^{2}$ read,%
\begin{equation}
\frac{d^{2}J^{1}}{d\tau^{2}}+2\Lambda_{2D}\frac{dJ^{1}}{d\tau}-8\Lambda
_{2D}\exp\left(  -\Lambda_{2D}\tau\right)  \frac{dJ^{2}}{d\tau}-4\Lambda
_{2D}^{2}\exp\left(  -2\Lambda_{2D}\tau\right)  J^{1}-8\Lambda_{2D}^{2}%
\exp\left(  -\Lambda_{2D}\tau\right)  J^{2}=0\text{,} \label{vit1}%
\end{equation}
and,%
\begin{equation}
\frac{d^{2}J^{2}}{d\tau^{2}}+2\Lambda_{2D}\exp\left(  -\Lambda_{2D}%
\tau\right)  \frac{dJ^{1}}{d\tau}+2\Lambda_{2D}\frac{dJ^{2}}{d\tau}%
+\Lambda_{2D}^{2}J^{2}=0\text{,} \label{vit2}%
\end{equation}
respectively. Following the line of reasoning provided for the
three-dimensional case, the asymptotic integration of (\ref{vit1}) and
(\ref{vit2}) lead to%
\begin{equation}
J^{1}\left(  \tau\right)  =C_{1}^{\left(  1\right)  }+C_{2}^{\left(  1\right)
}\exp\left(  -2\Lambda_{2D}\tau\right)  \text{,} \label{1j}%
\end{equation}
and,%
\begin{equation}
J^{2}\left(  \tau\right)  =C_{1}^{\left(  2\right)  }\exp\left(  -\Lambda
_{2D}\tau\right)  +C_{2}^{\left(  2\right)  }\tau\exp\left(  -\Lambda_{2D}%
\tau\right)  \text{,} \label{2j}%
\end{equation}
respectively where $C_{k^{\prime}}^{\left(  k\right)  }$ with $k=1$, $2$ and
$k^{\prime}=1$, $2$ are \emph{real} integration constants. The Jacobi vector
field intensity $J_{\mathcal{M}_{s}}^{\left(  2D\right)  }\left(  \tau\right)
$ is defined as,%
\begin{equation}
J_{\mathcal{M}_{s}}^{\left(  2D\right)  }\left(  \tau\right)  \overset
{\text{def}}{=}\left[  \frac{\left[  J^{1}\left(  \tau\right)  \right]  ^{2}%
}{\sigma^{2}\left(  \tau\right)  }+\frac{4\left[  J^{2}\left(  \tau\right)
\right]  ^{2}}{\sigma^{2}\left(  \tau\right)  }\right]  ^{\frac{1}{2}}\text{.}%
\end{equation}
Eqs. (\ref{cac4}), (\ref{1j})\ and (\ref{2j}) imply that,%
\begin{equation}
\frac{\left[  J^{1}\left(  \tau\right)  \right]  ^{2}}{\sigma^{2}\left(
\tau\right)  }\approx\frac{\left[  C_{1}^{\left(  1\right)  }\right]  ^{2}%
}{4\sigma_{0}^{2}}\exp\left(  2\Lambda_{2D}\tau\right)  \text{, }%
\frac{4\left[  J^{2}\left(  \tau\right)  \right]  ^{2}}{\sigma^{2}\left(
\tau\right)  }\approx\frac{\left[  C_{2}^{\left(  2\right)  }\right]  ^{2}%
}{\sigma_{0}^{2}}\tau^{2}\text{.}%
\end{equation}
It then follows that,%
\begin{equation}
J_{\mathcal{M}_{s}}^{\left(  2D\right)  }\left(  \tau\right)  \approx
\frac{C_{1}^{\left(  1\right)  }}{2\sigma_{0}}\exp\left(  \Lambda_{2D}%
\tau\right)  \text{.}%
\end{equation}
Thus, we have shown that in the asymptotic limit,%
\begin{equation}
J_{\mathcal{M}_{s}}^{\left(  3D\right)  }\left(  \tau\right)  \approx
\frac{C_{1}^{\left(  1\right)  }}{2\sigma_{0}}\exp\left(  \Lambda_{3D}%
\tau\right)  \text{ and, }J_{\mathcal{M}_{s}}^{\left(  2D\right)  }\left(
\tau\right)  \approx\frac{C_{1}^{\left(  1\right)  }}{2\sigma_{0}}\exp\left(
\Lambda_{2D}\tau\right)  \text{, }%
\end{equation}
that is,%
\begin{equation}
J_{\mathcal{M}_{s}}^{\left(  2D\right)  }\left(  \tau\right)  \approx
e^{-\left(  \Lambda_{3D}-\Lambda_{2D}\right)  \tau}\cdot J_{\mathcal{M}_{s}%
}^{\left(  3D\right)  }\left(  \tau\right)  \text{ with }\Lambda_{3D}%
-\Lambda_{2D}>0\text{. } \label{Jfinal}%
\end{equation}
Eq. (\ref{Jfinal}) is quite enlightening since it shows that the Jacobi vector
field intensity is softened when approaching the two-dimensional case from the
three-dimensional case via the introduction of the quantum-like macroscopic
constraint (\ref{MC}).

We emphasize that our main findings (see Eqs. (\ref{IGEfinal}) and
(\ref{Jfinal})) presented in this work are limited to the asymptotic behavior
at infinity on the $\tau$-scale of the selected indicators of chaoticity, that
is, the IGE and the JLC vector field intensity. However, recall that in
quantum chaos, the shortest random time scale (or Ehrenfest time scale)
$t_{r}$ is approximately given by \cite{zas},%

\begin{equation}
t_{r}\approx\frac{1}{\lambda}\log\left(  \frac{S}{\hbar}\right)  \text{,}%
\end{equation}
where\textbf{ }$\lambda$ is the maximum Lyapunov exponent of the system,
$\hbar\overset{\text{def}}{=}\frac{h}{2\pi}$ is the Dirac constant, $h$ is the
Planck constant and $S\simeq\int pdq$ is a characteristic reference value of a
classical action. This time scale is especially important because on time
scales of this order, the complete transition from quantum to classical
dynamical chaos is allowed. Stated otherwise, on time scales of this order,
quantum motion is similar to the classical one including the exponential
instability \cite{casa}. Therefore, it would be worthwhile investigating
whether or not for the statistical dynamical Gaussian models studied here,
there is any information geometric analogue of the standard random time scale,
say\textbf{ }$\tau_{r}$\textbf{, }such that one may find softening effects on
a $\tau$-scale longer than a finite $\tau_{r}$. This investigation would
present two delicate points: first, we would need the JLC-equation analysis on
finite-$\tau$ scales which would be extremely difficult from a computational
point of view; second, our geodesic affine parameter $\tau$ is not the
conventional time $t$. Therefore, unless the connection between $\tau$ and $t$
is clearly specified, this analogy would be admittedly vague. However, we
believe that this investigation may be successfully tackled within the IGAC
framework at least in specific cases, for instance for conservative chaotic
Hamiltonian systems. In such cases, there is a neat connection between the
standard time-scale and the geodesic parameter-scale and the difficulties in
the integration of the JLC equation may not be insurmountable
\cite{caticha-cafaro}. We will examine this issue in forthcoming
works\textbf{.}

\section{Concluding Remarks}

In this article, we studied both the information geometry and the entropic
dynamics of a $3D$ Gaussian statistical model. We then compared our analysis
to that of a $2D$ Gaussian statistical model obtained from the
higher-dimensional model via introduction of an additional information
constraint that resembles the quantum mechanical canonical minimum uncertainty
relation. We showed that the chaoticity (temporal complexity) of the $2D$
Gaussian statistical model, quantified by means of the Information Geometric
Entropy (IGE) and the Jacobi vector field intensity, is softened with respect
to the chaoticity of the $3D$ Gaussian statistical model. Specifically, Eq.
(\ref{IGEfinal}) shows that the IGE\ is softened when approaching the
two-dimensional case from the three-dimensional case via the introduction of
the macroscopic constraint (\ref{MC}) that resembles the quantum mechanical
canonical minimum uncertainty relation. Furthermore, Eq. (\ref{Jfinal})
confirms that also the Jacobi vector field intensity is softened when
approaching the two-dimensional case from the three-dimensional case via the
introduction of the macroscopic constraint (\ref{MC}).

We stress that our information geometric analysis could be further generalized
to accommodate non minimum uncertainty-like relations. However, such extension
requires a more delicate analysis where Maximum relative Entropy methods are
used to process information in the presence of inequality constraints
\cite{ishwar}\textbf{.} Our work is especially relevant for the quantification
of\textbf{\ }soft chaos effects in entropic dynamical models used to describe
actual physical systems when only incomplete knowledge about them is available
\cite{AM}. Furthermore, although we are aware that our analysis is not
manifestly "quantum", our findings lead us to support the conjecture that
quantum chaos is ultimately weaker than classical chaos because of
Heisenberg's uncertainty relation, the most important difference between
classical and quantum physics. Of course, a deeper analysis is needed and we
leave it for future investigations.

\begin{acknowledgments}
The research of C. Cafaro, C. Lupo and S. Mancini has received funding from
the European Commission's Seventh Framework Programme (FP7/2007--2013) under
grant agreements no. 213681.
\end{acknowledgments}

\appendix

\section{Integration of Geodesic Equations}

We show here the details leading to the geodesic paths presented in Section III.

Consider the following set of coupled nonlinear differential equations,%
\begin{equation}
\frac{d^{2}\mu_{x}}{d\tau^{2}}-\frac{2}{\sigma_{x}}\frac{d\mu_{x}}{d\tau}%
\frac{d\sigma_{x}}{d\tau}=0\text{ and, }\frac{d^{2}\sigma_{x}}{d\tau^{2}%
}+\frac{1}{2\sigma_{x}}\left(  \frac{d\mu_{x}}{d\tau}\right)  ^{2}-\frac
{1}{\sigma_{x}}\left(  \frac{d\sigma_{x}}{d\tau}\right)  ^{2}=0\text{.}
\label{A1}%
\end{equation}
Setting $\dot{\mu}_{x}\overset{\text{def}}{=}\frac{d\mu_{x}}{d\tau}$ and
$\dot{\sigma}_{x}\overset{\text{def}}{=}\frac{d\sigma_{x}\left(  \tau\right)
}{d\tau}$, the two equations in (\ref{A1}) become,%
\begin{equation}
\ddot{\mu}_{x}-2\frac{\dot{\sigma}_{x}}{\sigma_{x}}\dot{\mu}_{x}=0\text{ and,
}\ddot{\sigma}_{x}+\frac{1}{2\sigma_{x}}\dot{\mu}_{x}^{2}-\frac{\dot{\sigma
}_{x}^{2}}{\sigma_{x}}=0\text{.} \label{a}%
\end{equation}
From (\ref{a}) we observe that%
\begin{equation}
\frac{\ddot{\mu}_{x}}{\dot{\mu}_{x}}=2\frac{\dot{\sigma}_{x}}{\sigma_{x}%
}\text{ .}%
\end{equation}
Therefore, we have that%
\begin{equation}
\dot{\mu}_{x}\left(  \tau\right)  =A_{1}\sigma_{x}^{2}\left(  \tau\right)
\text{,} \label{b}%
\end{equation}
with $A_{1}\in%
%TCIMACRO{\U{211d} }%
%BeginExpansion
\mathbb{R}
%EndExpansion
$. Substituting (\ref{b}) in the second equation in (\ref{a}), we get%
\begin{equation}
\text{ }\sigma_{x}\ddot{\sigma}_{x}-\dot{\sigma}_{x}^{2}+\frac{A_{1}^{2}}%
{2}\sigma_{x}^{4}=0\text{.} \label{c}%
\end{equation}
Our goal is to integrate Eq. (\ref{c}), find $\sigma_{x}\left(  \tau\right)  $
and finally compute $\mu_{x}\left(  \tau\right)  $ using (\ref{b}). For the
sake of simplicity, let us put $\sigma_{x}\left(  \tau\right)  \equiv y\left(
\tau\right)  $ and $a\overset{\text{def}}{=}\frac{A_{1}^{2}}{2}\in%
%TCIMACRO{\U{211d} }%
%BeginExpansion
\mathbb{R}
%EndExpansion
_{0}^{+}$. Then, Eq. (\ref{c}) becomes%
\begin{equation}
\text{ }y\ddot{y}-\dot{y}^{2}+ay^{4}=0\text{.} \label{d}%
\end{equation}
As a first change of variables, let us set%
\begin{equation}
y\left(  \tau\right)  \overset{\text{def}}{=}\frac{dx\left(  \tau\right)
}{d\tau}=\dot{x}\left(  \tau\right)  \text{.} \label{f}%
\end{equation}
Substituting (\ref{f}) into (\ref{d}), we obtain%
\begin{equation}
\dot{x}\dddot{x}-\ddot{x}^{2}+a\dot{x}^{4}=0\text{.} \label{q}%
\end{equation}
As a second change of variables, let us set%
\begin{equation}
\dot{x}=\frac{dx\left(  \tau\right)  }{d\tau}\overset{\text{def}}{=}z\left(
x\right)  \text{.} \label{g}%
\end{equation}
Therefore, it follows that%
\begin{equation}
\ddot{x}=zz^{\prime}\text{, }\dddot{x}=\left(  z^{\prime\prime}z+z^{\prime
2}\right)  z\text{, }z^{\prime}=\frac{dz\left(  x\right)  }{dx}\text{.}
\label{h}%
\end{equation}
Substituting (\ref{h}) and (\ref{g}) in (\ref{q}), we get%
\begin{equation}
z^{\prime\prime}+az=0\text{.} \label{1}%
\end{equation}
Integration of (\ref{1}) yields,%
\begin{equation}
z\left(  x\right)  =C_{1}\sin\left(  \sqrt{a}x+C_{2}\right)  \text{,}%
\end{equation}
where the integration constant coefficients $C_{1}$ and $C_{2}$ belong to $%
%TCIMACRO{\U{211d} }%
%BeginExpansion
\mathbb{R}
%EndExpansion
$. Recalling that $\dot{x}=z\left(  x\right)  $, we get%
\begin{equation}
\int^{x}\frac{1}{C_{1}\sin\left(  \sqrt{a}x^{\prime}+C_{2}\right)  }%
dx^{\prime}=\int^{\tau}d\tau^{\prime}\text{.} \label{a11}%
\end{equation}
Upon integration, (\ref{a11}) becomes%
\begin{equation}
\frac{1}{\sqrt{a}C_{1}}\log\left[  \tan\left(  \frac{\sqrt{a}x+C_{2}}%
{2}\right)  \right]  =\tau+C_{3}\text{,}%
\end{equation}
with $C_{3}\in%
%TCIMACRO{\U{211d} }%
%BeginExpansion
\mathbb{R}
%EndExpansion
$. Solving for $x=x\left(  \tau\right)  $, we finally obtain%
\begin{equation}
x\left(  \tau\right)  =\frac{1}{\sqrt{a}}\left\{  2\arctan\left(  \exp\left[
\sqrt{a}C_{1}\left(  \tau+C_{3}\right)  \right]  \right)  -C_{2}\right\}
\text{.}%
\end{equation}
Finally, $\sigma_{x}\left(  \tau\right)  $ reads%
\begin{equation}
\sigma_{x}\left(  \tau\right)  =\frac{2c_{1}\exp\left(  c_{1}\sqrt{c_{2}}%
\tau+c_{1}c_{3}\sqrt{c_{2}}\right)  }{1+\exp\left(  2c_{1}\sqrt{c_{2}}%
\tau+2c_{1}c_{3}\sqrt{c_{2}}\right)  }\text{,} \label{supp}%
\end{equation}
where we have set $c_{1}=C_{1}$, $c_{2}=a\overset{\text{def}}{=}\frac
{A_{1}^{2}}{2}$ and $c_{3}=C_{3}$. It is straightforward to verify that indeed
$\sigma_{x}\left(  \tau\right)  $ in (\ref{supp}) satisfies the nonlinear
differential equation (\ref{d}). Finally, we can compute $\mu_{x}\left(
\tau\right)  $ using (\ref{b}). It follows that,%
\begin{equation}
\mu_{x}\left(  \tau\right)  =\sqrt{4a}\int^{\tau}\sigma_{x}^{2}\left(
\tau^{\prime}\right)  d\tau^{\prime}=\frac{c_{4}\left[  1+\exp\left(
2c_{1}\sqrt{c_{2}}\tau+2c_{1}c_{3}\sqrt{c_{2}}\right)  \right]  -4c_{1}%
}{1+\exp\left(  2c_{1}\sqrt{c_{2}}\tau+2c_{1}c_{3}\sqrt{c_{2}}\right)
}\text{.}%
\end{equation}
Therefore, the geodesic paths are given by%
\begin{equation}
\mu_{x}\left(  \tau\right)  =\frac{c_{4}\left[  1+\exp\left(  2c_{1}%
\sqrt{c_{2}}\tau+2c_{1}c_{3}\sqrt{c_{2}}\right)  \right]  -4c_{1}}%
{1+\exp\left(  2c_{1}\sqrt{c_{2}}\tau+2c_{1}c_{3}\sqrt{c_{2}}\right)  }\text{
and, }\sigma_{x}\left(  \tau\right)  =\frac{2c_{1}\exp\left(  c_{1}\sqrt
{c_{2}}\tau+c_{1}c_{3}\sqrt{c_{2}}\right)  }{1+\exp\left(  2c_{1}\sqrt{c_{2}%
}\tau+2c_{1}c_{3}\sqrt{c_{2}}\right)  }\text{ .} \label{one}%
\end{equation}
As working hypothesis, we consider geodesic paths with $c_{3}=0$ and assume
that the initial conditions given by $\mu_{x}\left(  0\right)  =\mu_{0}$ and
$\sigma\left(  0\right)  =\sigma_{0}$. The geodesics in (\ref{one}) become,%
\begin{equation}
\mu_{x}\left(  \tau\right)  =\frac{\left(  \mu_{0}+2\sigma_{0}\right)  \left[
1+\exp\left(  2\sigma_{0}\lambda_{+}^{\prime}\tau\right)  \right]
-4\sigma_{0}}{1+\exp\left(  2\sigma_{0}\lambda_{+}^{\prime}\tau\right)
}\text{, }\sigma_{x}\left(  \tau\right)  =\frac{2\sigma_{0}\exp\left(
\sigma_{0}\lambda_{+}^{\prime}\tau\right)  }{1+\exp\left(  2\sigma_{0}%
\lambda_{+}^{\prime}\tau\right)  }\text{,}%
\end{equation}
where $\lambda_{+}^{\prime}=\sqrt{c_{2}}=\sqrt{a}>0$ since $a\overset
{\text{def}}{=}\frac{A_{1}^{2}}{2}>0$.

Using the same tricks (invertible changes of variables) presented above, it
becomes straightforward integrating the last differential equation
characterizing the three dimensional case,%
\begin{equation}
\sigma_{y}\ddot{\sigma}_{y}-\dot{\sigma}_{y}^{2}=0\text{.}%
\end{equation}
Its solution $\sigma_{y}\left(  \tau\right)  $ is given by,%
\begin{equation}
\sigma_{y}\left(  \tau\right)  =\sigma_{0}^{\prime}\exp\left(  -\lambda
_{f}\tau\right)  \text{,}%
\end{equation}
with $\sigma_{y}\left(  0\right)  =\sigma_{0}^{\prime}$, $\sigma_{y}\left(
\tau_{f}\right)  =\varepsilon$ and $\lambda_{f}\overset{\text{def}}{=}\frac
{1}{\tau_{f}}\log\left(  \frac{\sigma_{0}^{\prime}}{\varepsilon}\right)  $.

Finally, following the line of reasoning presented above and assuming the very
same working hypothesis and initial conditions, it turns out that the geodesic
paths in the two-dimensional case read,%
\begin{equation}
\mu_{x}\left(  \tau\right)  =\frac{\left(  \mu_{0}+2\sigma_{0}\right)  \left[
1+\exp\left(  2\sigma_{0}\lambda_{+}\tau\right)  \right]  -4\sigma_{0}}%
{1+\exp\left(  2\sigma_{0}\lambda_{+}\tau\right)  }\text{, }\sigma\left(
\tau\right)  =\frac{2\sigma_{0}\exp\left(  \sigma_{0}\lambda_{+}\tau\right)
}{1+\exp\left(  2\sigma_{0}\lambda_{+}\tau\right)  }\text{,}%
\end{equation}
where $\lambda_{+}=\sqrt{a}>0$ since $a\overset{\text{def}}{=}\frac{A_{1}^{2}%
}{4}>0$.

\section{Integration of Jacobi-Levi-Civita Equations}

We present here few technical details relative to the JLC equation analysis
introduced in Section IV.

The JLC equations in the three-dimensional case are three. The JLC equation
for $J^{1}$ reads,%
\begin{equation}
\frac{D^{2}J^{1}}{D\tau^{2}}-\frac{1}{\sigma_{x}^{2}}\left[  J^{1}\left(
\frac{d\sigma_{x}}{d\tau}\right)  ^{2}-J^{2}\frac{d\mu_{x}}{d\tau}%
\frac{d\sigma_{x}}{d\tau}\right]  =0\text{,} \label{J1}%
\end{equation}
where we have used the relation $R_{212}^{1}=-\frac{1}{\sigma_{x}^{2}}$.
Recall that,%
\begin{equation}
\frac{D^{2}J^{\mu}}{D\tau^{2}}=\frac{d^{2}J^{\mu}}{d\tau^{2}}+2\Gamma
_{\alpha\beta}^{\mu}\frac{dJ^{\alpha}}{d\tau}\frac{d\theta^{\beta}}{d\tau
}+\Gamma_{\alpha\beta}^{\mu}J^{\alpha}\frac{d^{2}\theta^{\beta}}{d\tau^{2}%
}+\partial_{\nu}\Gamma_{\alpha\beta}^{\mu}\frac{d\theta^{\nu}}{d\tau}%
\frac{d\theta^{\beta}}{d\tau}J^{\alpha}+\Gamma_{\alpha\beta}^{\mu}\Gamma
_{\rho\sigma}^{\alpha}\frac{d\theta^{\sigma}}{d\tau}\frac{d\theta^{\beta}%
}{d\tau}J^{\rho}\text{.}%
\end{equation}
For $J^{1}$ we get,%
\begin{align}
\frac{D^{2}J^{1}}{D\tau^{2}}  &  =\frac{d^{2}J^{1}}{d\tau^{2}}+2\Gamma
_{12}^{1}\left(  \frac{dJ^{1}}{d\tau}\frac{d\theta^{2}}{d\tau}+\frac{dJ^{2}%
}{d\tau}\frac{d\theta^{1}}{d\tau}\right)  +\Gamma_{12}^{1}\left(  J^{1}%
\frac{d^{2}\theta^{2}}{d\tau^{2}}+J^{2}\frac{d^{2}\theta^{1}}{d\tau^{2}%
}\right)  +\nonumber\\
& \nonumber\\
&  +\partial_{2}\Gamma_{12}^{1}\left[  \left(  \frac{d\theta^{2}}{d\tau
}\right)  ^{2}J^{1}+\frac{d\theta^{1}}{d\tau}\frac{d\theta^{2}}{d\tau}%
J^{2}\right]  +\left(  \Gamma_{12}^{1}\right)  ^{2}\left(  \frac{d\theta^{2}%
}{d\tau}\right)  ^{2}J^{1}+2\left(  \Gamma_{12}^{1}\right)  ^{2}\frac
{d\theta^{1}}{d\tau}\frac{d\theta^{2}}{d\tau}+\Gamma_{12}^{1}\Gamma_{11}%
^{2}\left(  \frac{d\theta^{1}}{d\tau}\right)  ^{2}J^{1}\text{,} \label{J2}%
\end{align}
where we have used the identity,%
\begin{equation}
\Gamma_{\alpha\beta}^{1}\Gamma_{\rho\sigma}^{\alpha}\frac{d\theta^{\sigma}%
}{d\tau}\frac{d\theta^{\beta}}{d\tau}J^{\rho}=\left(  \Gamma_{12}^{1}\right)
^{2}\left(  \frac{d\theta^{2}}{d\tau}\right)  ^{2}J^{1}+2\left(  \Gamma
_{12}^{1}\right)  ^{2}\frac{d\theta^{1}}{d\tau}\frac{d\theta^{2}}{d\tau
}+\Gamma_{12}^{1}\Gamma_{11}^{2}\left(  \frac{d\theta^{1}}{d\tau}\right)
^{2}J^{1}\text{.}%
\end{equation}
since $\Gamma_{12}^{1}=\Gamma_{21}^{1}=\Gamma_{22}^{2}$. From (\ref{J1})\ and
(\ref{J2}) and after some algebra, we get%
\begin{align}
0  &  =\frac{d^{2}J^{1}}{d\tau^{2}}+2\Gamma_{12}^{1}\frac{d\sigma_{x}}{d\tau
}\frac{dJ^{1}}{d\tau}+2\Gamma_{12}^{1}\frac{d\mu_{x}}{d\tau}\frac{dJ^{2}%
}{d\tau}+\nonumber\\
& \nonumber\\
&  +J^{1}\left[  \Gamma_{12}^{1}\frac{d^{2}\sigma_{x}}{d\tau^{2}}+\partial
_{2}\Gamma_{12}^{1}\left(  \frac{d\sigma_{x}}{d\tau}\right)  ^{2}+\left(
\Gamma_{12}^{1}\right)  ^{2}\left(  \frac{d\sigma_{x}}{d\tau}\right)
^{2}+\Gamma_{12}^{1}\Gamma_{11}^{2}\left(  \frac{d\mu_{x}}{d\tau}\right)
^{2}+R_{212}^{1}\left(  \frac{d\sigma_{x}}{d\tau}\right)  ^{2}\right]
+\nonumber\\
& \nonumber\\
&  +J^{2}\left[  \Gamma_{12}^{1}\frac{d^{2}\mu_{x}}{d\tau^{2}}+\partial
_{2}\Gamma_{12}^{1}\frac{d\mu_{x}}{d\tau}\frac{d\sigma_{x}}{d\tau}+2\left(
\Gamma_{12}^{1}\right)  ^{2}\frac{d\mu_{x}}{d\tau}\frac{d\sigma_{x}}{d\tau
}-R_{212}^{1}\frac{d\mu_{x}}{d\tau}\frac{d\sigma_{x}}{d\tau}\right]  \text{,}%
\end{align}
that is,%
\begin{align}
0  &  =\frac{d^{2}J^{1}}{d\tau^{2}}+2\Gamma_{12}^{1}\frac{d\sigma_{x}}{d\tau
}\frac{dJ^{1}}{d\tau}+2\Gamma_{12}^{1}\frac{d\mu_{x}}{d\tau}\frac{dJ^{2}%
}{d\tau}+\nonumber\\
& \nonumber\\
&  +J^{1}\left[  \Gamma_{12}^{1}\frac{d^{2}\sigma_{x}}{d\tau^{2}}+\left(
\partial_{2}\Gamma_{12}^{1}+\left(  \Gamma_{12}^{1}\right)  ^{2}+R_{212}%
^{1}\right)  \left(  \frac{d\sigma_{x}}{d\tau}\right)  ^{2}+\Gamma_{12}%
^{1}\Gamma_{11}^{2}\left(  \frac{d\mu_{x}}{d\tau}\right)  ^{2}\right]
+\nonumber\\
& \nonumber\\
&  +J^{2}\left[  \Gamma_{12}^{1}\frac{d^{2}\mu_{x}}{d\tau^{2}}+\left(
\partial_{2}\Gamma_{12}^{1}+2\left(  \Gamma_{12}^{1}\right)  ^{2}-R_{212}%
^{1}\right)  \frac{d\mu_{x}}{d\tau}\frac{d\sigma_{x}}{d\tau}\right]  \text{.}
\label{J3}%
\end{align}
Observing that,%
\begin{equation}
\left(  \Gamma_{12}^{1}\right)  ^{2}=\frac{1}{\sigma_{x}^{2}}\text{, }%
R_{212}^{1}=-\frac{1}{\sigma_{x}^{2}}\text{ and, }\partial_{2}\Gamma_{12}%
^{1}+2\left(  \Gamma_{12}^{1}\right)  ^{2}-R_{212}^{1}=\frac{4}{\sigma_{x}%
^{2}}\text{,}%
\end{equation}
Eq. (\ref{J3}) becomes,%
\begin{align}
0  &  =\frac{d^{2}J^{1}}{d\tau^{2}}+2\Gamma_{12}^{1}\frac{d\sigma_{x}}{d\tau
}\frac{dJ^{1}}{d\tau}+2\Gamma_{12}^{1}\frac{d\mu_{x}}{d\tau}\frac{dJ^{2}%
}{d\tau}+\nonumber\\
& \nonumber\\
&  +J^{1}\left[  \Gamma_{12}^{1}\frac{d^{2}\sigma_{x}}{d\tau^{2}}+\partial
_{2}\Gamma_{12}^{1}\left(  \frac{d\sigma_{x}}{d\tau}\right)  ^{2}+\Gamma
_{12}^{1}\Gamma_{11}^{2}\left(  \frac{d\mu_{x}}{d\tau}\right)  ^{2}\right]
+\nonumber\\
& \nonumber\\
&  +J^{2}\left[  \Gamma_{12}^{1}\frac{d^{2}\mu_{x}}{d\tau^{2}}+\frac{4}%
{\sigma_{x}^{2}}\frac{d\mu_{x}}{d\tau}\frac{d\sigma_{x}}{d\tau}\right]
\text{,}%
\end{align}
that is,%
\begin{align}
0  &  =\frac{d^{2}J^{1}}{d\tau^{2}}+\left(  -\frac{2}{\sigma_{x}}\frac
{d\sigma_{x}}{d\tau}\right)  \frac{dJ^{1}}{d\tau}+\left(  -\frac{2}{\sigma
_{x}}\frac{d\mu_{x}}{d\tau}\right)  \frac{dJ^{2}}{d\tau}+\nonumber\\
& \nonumber\\
&  +J^{1}\left[  -\frac{1}{\sigma_{x}}\frac{d^{2}\sigma_{x}}{d\tau^{2}}%
+\frac{1}{\sigma_{x}^{2}}\left(  \frac{d\sigma_{x}}{d\tau}\right)  ^{2}%
-\frac{1}{2\sigma_{x}^{2}}\left(  \frac{d\mu_{x}}{d\tau}\right)  ^{2}\right]
+\nonumber\\
& \nonumber\\
&  +J^{2}\left[  -\frac{1}{\sigma_{x}}\frac{d^{2}\mu_{x}}{d\tau^{2}}+\frac
{4}{\sigma_{x}^{2}}\frac{d\mu_{x}}{d\tau}\frac{d\sigma_{x}}{d\tau}\right]
\text{.} \label{J4}%
\end{align}
At this point we recall that the geodesic paths $\mu_{x}\left(  \tau\right)  $
and $\sigma_{x}\left(  \tau\right)  $ for the $3D$ statistical model are given
by,%
\begin{equation}
\mu_{x}\left(  \tau\right)  =\frac{\left(  \mu_{0}+2\sigma_{0}\right)  \left[
1+\exp\left(  2\sqrt{a}\sigma_{0}\tau\right)  \right]  -4\sigma_{0}}%
{1+\exp\left(  2\sqrt{a}\sigma_{0}\tau\right)  }\text{, }%
\end{equation}
and,%
\begin{equation}
\sigma_{x}\left(  \tau\right)  =\frac{2\sigma_{0}\exp\left(  \sqrt{a}%
\sigma_{0}\tau\right)  }{1+\exp\left(  2\sqrt{a}\sigma_{0}\tau\right)
}\text{,}%
\end{equation}
respectively. Therefore, in the asymptotic limit we get%
\begin{align}
\mu_{x}\left(  \tau\right)   &  \approx\text{const., }\sigma_{x}\left(
\tau\right)  \approx2\sigma_{0}\exp\left(  -\sqrt{a}\sigma_{0}\tau\right)
\text{,}\nonumber\\
& \nonumber\\
\frac{d\sigma_{x}\left(  \tau\right)  }{d\tau}  &  \approx-2\sqrt{a}\sigma
_{0}^{2}\exp\left(  -\sqrt{a}\sigma_{0}\tau\right)  \text{, }\left(
\frac{d\sigma_{x}\left(  \tau\right)  }{d\tau}\right)  ^{2}\approx4a\sigma
_{0}^{4}\exp(-2\sqrt{a}\sigma_{0}\tau)\text{,}\nonumber\\
& \nonumber\\
\frac{d\mu_{x}\left(  \tau\right)  }{d\tau}  &  \approx8\sqrt{a}\sigma_{0}%
^{2}\exp\left(  -2\sqrt{a}\sigma_{0}\tau\right)  \text{, }\left(  \frac
{d\mu_{x}\left(  \tau\right)  }{d\tau}\right)  ^{2}\approx64a\sigma_{0}%
^{4}\exp\left(  -4\sqrt{a}\sigma_{0}\tau\right)  \text{,}\nonumber\\
&  \text{ }\nonumber\\
\frac{d^{2}\mu_{x}\left(  \tau\right)  }{d\tau^{2}}  &  \approx-16a\sigma
_{0}^{3}\exp\left(  -2\sqrt{a}\sigma_{0}\tau\right)  \text{, }\frac
{d^{2}\sigma_{x}\left(  \tau\right)  }{d\tau^{2}}\approx2a\sigma_{0}^{3}%
\exp\left(  -\sqrt{a}\sigma_{0}\tau\right)  \text{.} \label{asym}%
\end{align}
It then follows that,%
\begin{equation}
-\frac{2}{\sigma_{x}}\frac{d\sigma_{x}}{d\tau}\approx2\sqrt{a}\sigma
_{0}\text{, }-\frac{2}{\sigma_{x}}\frac{d\mu_{x}}{d\tau}\approx-8\sqrt
{a}\sigma_{0}\exp\left(  -\sqrt{a}\sigma_{0}\tau\right)  \text{,}%
\end{equation}
and,%
\begin{align}
-\frac{1}{\sigma_{x}}\frac{d^{2}\mu_{x}\left(  \tau\right)  }{d\tau^{2}}%
+\frac{4}{\sigma_{x}^{2}}\frac{d\mu_{x}\left(  \tau\right)  }{d\tau}%
\frac{d\sigma_{x}\left(  \tau\right)  }{d\tau}  &  \approx-8a\sigma_{0}%
^{2}\exp\left(  -\sqrt{a}\sigma_{0}\tau\right)  \text{,}\nonumber\\
& \nonumber\\
-\frac{1}{\sigma_{x}}\frac{d^{2}\sigma_{x}\left(  \tau\right)  }{d\tau^{2}%
}+\frac{1}{\sigma_{x}^{2}}\left(  \frac{d\sigma_{x}\left(  \tau\right)
}{d\tau}\right)  ^{2}-\frac{1}{2\sigma_{x}^{2}}\left(  \frac{d\mu_{x}\left(
\tau\right)  }{d\tau}\right)  ^{2}  &  \approx-8a\sigma_{0}^{2}\exp(-2\sqrt
{a}\sigma_{0}\tau)\text{.}%
\end{align}
Therefore, setting $\Lambda_{3D}\overset{\text{def}}{=}\lambda_{3D}\sigma_{0}$
with $\lambda_{3D}=\lambda_{+}^{\prime}$ , the JLC equation for $J^{1}$
becomes,%
\begin{equation}
\frac{d^{2}J^{1}}{d\tau^{2}}+2\Lambda_{3D}\frac{dJ^{1}}{d\tau}-8\Lambda
_{3D}\exp\left(  -\Lambda_{3D}\tau\right)  \frac{dJ^{2}}{d\tau}-8\Lambda
_{3D}^{2}\exp\left(  -2\Lambda_{3D}\tau\right)  J^{1}-8\Lambda_{3D}^{2}%
\exp\left(  -\Lambda_{3D}\tau\right)  J^{2}=0\text{.}%
\end{equation}
For $J^{2}$ we get,%
\begin{equation}
\frac{D^{2}J^{2}}{D\tau^{2}}-\frac{1}{2\sigma_{x}^{2}}\left[  J^{2}\left(
\frac{d\mu_{x}}{d\tau}\right)  ^{2}-J^{1}\frac{d\mu_{x}}{d\tau}\frac
{d\sigma_{x}}{d\tau}\right]  =0\text{.} \label{J6}%
\end{equation}
where we used the identity $R_{121}^{2}=-\frac{1}{2\sigma_{x}^{2}}$. Notice
that,%
\begin{align}
\frac{D^{2}J^{2}}{D\tau^{2}}  &  =\frac{d^{2}J^{2}}{d\tau^{2}}+2\left(
\Gamma_{11}^{2}\frac{dJ^{1}}{d\tau}\frac{d\theta^{1}}{d\tau}+\Gamma_{22}%
^{2}\frac{dJ^{2}}{d\tau}\frac{d\theta^{2}}{d\tau}\right)  +\left(  \Gamma
_{11}^{2}J^{1}\frac{d^{2}\theta^{1}}{d\tau^{2}}+\Gamma_{22}^{2}J^{2}%
\frac{d^{2}\theta^{2}}{d\tau^{2}}\right)  +\nonumber\\
& \nonumber\\
&  +\left[  \partial_{2}\Gamma_{11}^{2}\frac{d\theta^{1}}{d\tau}\frac
{d\theta^{2}}{d\tau}J^{1}+\partial_{2}\Gamma_{22}^{2}\left(  \frac{d\theta
^{2}}{d\tau}\right)  ^{2}J^{2}\right]  +\nonumber\\
& \nonumber\\
&  +\left[  \Gamma_{11}^{2}\Gamma_{12}^{1}\frac{d\theta^{1}}{d\tau}%
\frac{d\theta^{2}}{d\tau}J^{1}+\Gamma_{11}^{2}\Gamma_{21}^{1}\left(
\frac{d\theta^{1}}{d\tau}\right)  ^{2}J^{2}+\Gamma_{22}^{2}\Gamma_{11}%
^{2}\frac{d\theta^{1}}{d\tau}\frac{d\theta^{2}}{d\tau}J^{1}+\left(
\Gamma_{22}^{2}\right)  ^{2}\left(  \frac{d\theta^{2}}{d\tau}\right)
^{2}J^{2}\right]  \text{.} \label{J7}%
\end{align}
From (\ref{J6}) and (\ref{J7}), the JLC equation for $J^{2}$ becomes after
some algebra,%
\begin{align}
0  &  =\frac{d^{2}J^{2}}{d\tau^{2}}+2\left(  \Gamma_{11}^{2}\frac{dJ^{1}%
}{d\tau}\frac{d\theta^{1}}{d\tau}+\Gamma_{22}^{2}\frac{dJ^{2}}{d\tau}%
\frac{d\theta^{2}}{d\tau}\right)  +\nonumber\\
& \nonumber\\
&  +J^{1}\left[  \Gamma_{11}^{2}\frac{d^{2}\theta^{1}}{d\tau^{2}}+\partial
_{2}\Gamma_{11}^{2}\frac{d\theta^{1}}{d\tau}\frac{d\theta^{2}}{d\tau}%
+\Gamma_{11}^{2}\Gamma_{12}^{1}\frac{d\theta^{1}}{d\tau}\frac{d\theta^{2}%
}{d\tau}+\Gamma_{22}^{2}\Gamma_{11}^{2}\frac{d\theta^{1}}{d\tau}\frac
{d\theta^{2}}{d\tau}-R_{121}^{2}\frac{d\theta^{1}}{d\tau}\frac{d\theta^{2}%
}{d\tau}\right]  +\nonumber\\
& \nonumber\\
&  +J^{2}\left[  \Gamma_{22}^{2}\frac{d^{2}\theta^{2}}{d\tau^{2}}+\partial
_{2}\Gamma_{22}^{2}\left(  \frac{d\theta^{2}}{d\tau}\right)  ^{2}+\Gamma
_{11}^{2}\Gamma_{21}^{1}\left(  \frac{d\theta^{1}}{d\tau}\right)  ^{2}+\left(
\Gamma_{22}^{2}\right)  ^{2}\left(  \frac{d\theta^{2}}{d\tau}\right)
^{2}+R_{121}^{2}\left(  \frac{d\theta^{1}}{d\tau}\right)  ^{2}\right]
\text{.}%
\end{align}
Recalling that $\theta^{1}=\mu_{x}$ and $\theta^{2}=\sigma_{x}$, we get%
\begin{align}
0  &  =\frac{d^{2}J^{2}}{d\tau^{2}}+2\Gamma_{11}^{2}\frac{d\mu_{x}}{d\tau
}\frac{dJ^{1}}{d\tau}+2\Gamma_{22}^{2}\frac{d\sigma_{x}}{d\tau}\frac{dJ^{2}%
}{d\tau}+\nonumber\\
& \nonumber\\
&  +J^{1}\left[  \Gamma_{11}^{2}\frac{d^{2}\mu_{x}}{d\tau^{2}}+\left(
\partial_{2}\Gamma_{11}^{2}+\Gamma_{11}^{2}\Gamma_{12}^{1}+\Gamma_{22}%
^{2}\Gamma_{11}^{2}-R_{121}^{2}\right)  \frac{d\mu_{x}}{d\tau}\frac
{d\sigma_{x}}{d\tau}\right]  +\nonumber\\
& \nonumber\\
&  +J^{2}\left[  \Gamma_{22}^{2}\frac{d^{2}\sigma_{x}}{d\tau^{2}}+\left(
\partial_{2}\Gamma_{22}^{2}+\left(  \Gamma_{22}^{2}\right)  ^{2}\right)
\left(  \frac{d\sigma_{x}}{d\tau}\right)  ^{2}+\left(  \Gamma_{11}^{2}%
\Gamma_{21}^{1}+R_{121}^{2}\right)  \left(  \frac{d\mu_{x}}{d\tau}\right)
^{2}\right]  \text{.} \label{fucks}%
\end{align}
Noting that the following relations hold,%
\begin{equation}
\partial_{2}\Gamma_{11}^{2}+\Gamma_{11}^{2}\Gamma_{12}^{1}+\Gamma_{22}%
^{2}\Gamma_{11}^{2}-R_{121}^{2}=-\frac{1}{\sigma_{x}^{2}}\text{, }\partial
_{2}\Gamma_{22}^{2}+\left(  \Gamma_{22}^{2}\right)  ^{2}=\frac{2}{\sigma
_{x}^{2}}\text{ and, }\Gamma_{11}^{2}\Gamma_{21}^{1}+R_{121}^{2}=-\frac
{1}{\sigma_{x}^{2}}\text{,}%
\end{equation}
Eq. (\ref{fucks}) becomes,%
\begin{align}
0  &  =\frac{d^{2}J^{2}}{d\tau^{2}}+\left(  \frac{1}{\sigma_{x}}\frac{d\mu
_{x}}{d\tau}\right)  \frac{dJ^{1}}{d\tau}+\left(  -\frac{2}{\sigma_{x}}%
\frac{d\sigma_{x}}{d\tau}\right)  \frac{dJ^{2}}{d\tau}+\nonumber\\
& \nonumber\\
&  +J^{1}\left[  \frac{1}{2\sigma_{x}}\frac{d^{2}\mu_{x}}{d\tau^{2}}-\frac
{1}{\sigma_{x}^{2}}\frac{d\mu_{x}}{d\tau}\frac{d\sigma_{x}}{d\tau}\right]
+\nonumber\\
& \nonumber\\
&  +J^{2}\left[  -\frac{1}{\sigma_{x}}\frac{d^{2}\sigma_{x}}{d\tau^{2}}%
+\frac{2}{\sigma_{x}^{2}}\left(  \frac{d\sigma_{x}}{d\tau}\right)  ^{2}%
-\frac{1}{\sigma_{x}^{2}}\left(  \frac{d\mu_{x}}{d\tau}\right)  ^{2}\right]
\text{.} \label{J8}%
\end{align}
From (\ref{asym}) we get,%
\begin{equation}
\frac{1}{\sigma_{x}}\frac{d\mu_{x}}{d\tau}\approx4\sqrt{a}\sigma_{0}%
\exp\left(  -\sqrt{a}\sigma_{0}\tau\right)  \text{, }-\frac{2}{\sigma_{x}%
}\frac{d\sigma_{x}}{d\tau}\approx2\sqrt{a}\sigma_{0}\text{,} \label{a1}%
\end{equation}
and,%
\begin{align}
\frac{1}{2\sigma_{x}}\frac{d^{2}\mu_{x}\left(  \tau\right)  }{d\tau^{2}}%
-\frac{1}{\sigma_{x}^{2}}\frac{d\mu_{x}\left(  \tau\right)  }{d\tau}%
\frac{d\sigma_{x}\left(  \tau\right)  }{d\tau}  &  \approx0\text{,
}\nonumber\\
& \nonumber\\
-\frac{1}{\sigma_{x}}\frac{d^{2}\sigma_{x}\left(  \tau\right)  }{d\tau^{2}%
}+\frac{2}{\sigma_{x}^{2}}\left(  \frac{d\sigma_{x}\left(  \tau\right)
}{d\tau}\right)  ^{2}-\frac{1}{\sigma_{x}^{2}}\left(  \frac{d\mu_{x}\left(
\tau\right)  }{d\tau}\right)  ^{2}  &  \approx a\sigma_{0}^{2}\text{. }
\label{a2}%
\end{align}
Substituting (\ref{a1}) and (\ref{a2}) into (\ref{J8}), we obtain%
\begin{equation}
\frac{d^{2}J^{2}}{d\tau^{2}}+4\Lambda_{3D}\exp\left(  -\Lambda_{3D}%
\tau\right)  \frac{dJ^{1}}{d\tau}+2\Lambda_{3D}\frac{dJ^{2}}{d\tau}%
+\Lambda_{3D}^{2}J^{2}=0\text{.} \label{J9}%
\end{equation}
Finally, the JLC equation for $J^{3}$ is given by,%
\begin{equation}
\frac{D^{2}J^{3}}{D\tau^{2}}=0\text{.}%
\end{equation}
Notice that,%
\begin{equation}
\frac{D^{2}J^{3}}{D\tau^{2}}=\frac{d^{2}J^{3}}{d\tau^{2}}+\left(  2\Gamma
_{33}^{3}\frac{d\theta^{3}}{d\tau}\right)  \frac{dJ^{3}}{d\tau}+J^{3}\left[
\Gamma_{33}^{3}\frac{d^{2}\theta^{3}}{d\tau^{2}}+\left(  \partial_{3}%
\Gamma_{33}^{3}+\left(  \Gamma_{33}^{3}\right)  ^{2}\right)  \left(
\frac{d\theta^{3}}{d\tau}\right)  ^{2}\right]  \text{.}%
\end{equation}
Therefore, the JLC\ equation for $J^{3}$ becomes,%
\begin{equation}
\frac{d^{2}J^{3}}{d\tau^{2}}+\left(  -\frac{2}{\sigma_{y}}\frac{d\sigma_{y}%
}{d\tau}\right)  \frac{dJ^{3}}{d\tau}+J^{3}\left[  -\frac{1}{\sigma_{y}}%
\frac{d^{2}\sigma_{y}}{d\tau^{2}}+\frac{2}{\sigma_{y}^{2}}\left(
\frac{d\sigma_{y}}{d\tau}\right)  ^{2}\right]  =0\text{,} \label{J10}%
\end{equation}
having recalled that,%
\begin{equation}
\Gamma_{33}^{3}=-\frac{1}{\sigma_{y}}\text{ and, }\partial_{3}\Gamma_{33}%
^{3}+\left(  \Gamma_{33}^{3}\right)  ^{2}=\frac{2}{\sigma_{y}^{2}}\text{.}%
\end{equation}
Observing that,%
\begin{align}
\sigma_{y}\left(  \tau\right)   &  =\sigma_{0}^{\prime}\exp\left(
-\lambda_{f}\tau\right)  \text{, }\frac{d\sigma_{y}\left(  \tau\right)
}{d\tau}=-\sigma_{0}^{\prime}\lambda_{f}e^{-\tau\lambda_{f}}\text{, }\left(
\frac{d\sigma_{y}\left(  \tau\right)  }{d\tau}\right)  ^{2}=\sigma_{0}%
^{\prime2}\lambda_{f}^{2}e^{-2\tau\lambda_{f}}\text{,}\nonumber\\
& \nonumber\\
\frac{d^{2}\sigma_{y}\left(  \tau\right)  }{d\tau^{2}}  &  =\sigma_{0}%
^{\prime}\lambda_{f}^{2}e^{-\tau\lambda_{f}}\text{, }-\frac{2}{\sigma_{y}%
}\frac{d\sigma_{y}\left(  \tau\right)  }{d\tau}=2\lambda_{f}\text{, }-\frac
{1}{\sigma_{y}}\frac{d^{2}\sigma_{y}\left(  \tau\right)  }{d\tau^{2}}+\frac
{2}{\sigma_{y}^{2}}\left(  \frac{d\sigma_{y}\left(  \tau\right)  }{d\tau
}\right)  ^{2}=\lambda_{f}^{2}\text{,}%
\end{align}
the JLC equation for $J^{3}$ becomes,%
\begin{equation}
\frac{d^{2}J^{3}}{d\tau^{2}}+2\lambda_{f}\frac{dJ^{3}}{d\tau}+\lambda_{f}%
^{2}J^{3}=0\text{.}%
\end{equation}
Following the same line of reasoning, we are capable of computing (and
integrating) also the JLC\ equations for the $2D$ statistical model.

\end{document}